# Change energy photons of radiation, stimulating a photoluminescence in glasses and optical fiber, activated by bismuth.

Valeriy E. Ogluzdin

**Abstract.**

In the offered review ordering received and published by domestic and foreign researchers of the experimental results showing the phenomenon of a photoluminescence in glasses and optical fiber, activated by bismuth is executed, and from uniform positions representations about the process responsible for a photoluminescence in case of use for excitation of this environment of various laser sources are considered. At interpretation of process of a photoluminescence the known model considering mirror symmetry of features of frequency spectra of a photoluminescence (in this case the maximum values is used: humps or peaks of spectra of a photoluminescence) and the spectra characterising optical losses (absorption) of glass, activated by (atomic) bismuth. For the analysis values of lines of the bismuth, published in reference media are used. This model is added by earlier published specification of the author, according to which to a point of mirror symmetry of such spectra there corresponds frequency of radiation of the source raising a photoluminescence – mean arithmetical of the frequencies of peaks of spectra of a photoluminescence and frequencies of lines atomic bismuth.

# Change energy photons of radiation, stimulating a photoluminescence in glasses and optical fiber, activated by bismuth.
Ogluzdin V.E.

# «МЕХАНИЗМ ИЗМЕНЕНИЯ ЭНЕРГИИ ФОТОНОВ СВЕТОВОГО ИЗЛУЧЕНИЯ, ВОЗБУЖДАЮЩЕГО ФОТОЛЮМИНЕСЦЕНЦИЮ В АКТИВИРОВАННЫХ ВИСМУТОМ СТЕКЛЕ И ВОЛОКОННЫХ СВЕТОВОДАХ»

В. Е. Оглуздин

Аннотация.

В предлагаемом обзоре выполнена систематизация полученных и опубликованных отечественными и зарубежными исследователями экспериментальных результатов, демонстрирующих явление фотолюминесценции в стеклах и волоконных световодах, активированных висмутом, и с единых позиций рассмотрены представления о процессе, ответственном за фотолюминесценцию в случае использования для возбуждения этой среды различных лазерных источников. При интерпретации процесса фотолюминесценции использована известная модель, учитывающая зеркальную симметрию особенностей частотных спектров фотолюминесценции (в данном случае максимальных значений: горбов или пиков спектров фотолюминесценции) и линейчатыми спектрами, характеризующими оптические потери (поглощение) стекла, активированного висмутом. Для анализа использованы значения линий висмута, опубликованные в широкодоступных справочных изданиях. Эта модель дополнена ранее опубликованным уточнением автора, согласно которому точке зеркальной симметрии таких спектров соответствует частота излучения источника, возбуждающего фотолюминесценцию.

**Abstract.**

In the offered review ordering received and published by domestic and foreign researchers of the experimental results showing the phenomenon of a photoluminescence in glasses and optical fiber, activated by bismuth is executed, and from uniform positions representations about the process responsible for a photoluminescence in case of use for excitation of this environment of various laser sources are considered. At interpretation of process of a photoluminescence the known model considering mirror symmetry of features of frequency spectra of a photoluminescence (in this case the maximum values is used: humps or peaks of spectra of a photoluminescence) and the spectra characterising optical losses (absorption) of glass, activated by (atomic) bismuth. For the analysis values of lines of the bismuth, published in reference media are used. This model is added by earlier published specification of the author, according to which to a point of mirror symmetry of such spectra there corresponds frequency of radiation of the source raising a photoluminescence – mean arithmetical of the frequencies of peaks of spectra of a photoluminescence and frequencies of lines atomic bismuth.

*Ключевые слова: фотолюминесценция, волоконные световоды, висмутовый волоконный лазер.*
**Keywords:** photoluminescence, optical fiber, optical fiber bismuth laser.

Содержание:

4. Анализ опубликованных экспериментальных результатов по фотолюминесценции волоконных световодов, активированных висмутом и возбуждаемых с помощью источников лазерного излучения с различными длинами волн.
5. Заключение.

**The maintenance:**

1. Introduction.
2. Expected difference of results of optical experiments on supervision of the photoluminescence, executed with optical fiber activated by bismuth use, from results of the experiment executed under the traditional scheme with use of flat glass samples activated by bismuth.

3. A role of the law of conservation of energy in the model used for the description of features of a spectrum of a photoluminescence of a glass matrix activated by bismuth, considered within the limits of two-level model. Photoluminescence, Bohr correspondence principle and dispersion law of Lorenz classical harmonic oscillator.
4. The analysis of the published experimental results on a photoluminescence of the optical fiber and flat glass samples activated by bismuth and excited by means of sources laser radiation with various lengths of waves.
5. Conclusion.

1.Введение.

В последнее время появилось значительное количество работ, посвященных описанию явления фотолюминесценции в стеклах и волоконных световодах, активированных висмутом. Интерес исследователей к данной проблеме обусловлен тем обстоятельством, что ближняя инфракрасная область спектра представляет особый интерес для специалистов, занятых получением лазерной генерации в области 1000 – 1500 мкм, а также – актуальной для данной области спектра возможности передачи по стекловолокну значительных объемов информации.

В предлагаемом обзоре выполнена систематизация ряда полученных и опубликованных отечественными и зарубежными исследователями экспериментальных результатов, демонстрирующих это явление, и с единых позиций рассмотрены представления о процессе, ответственном за фотолюминесценцию в случае использования различных лазерных источников для возбуждения среды, активированной висмутом.

При интерпретации процесса фотолюминесценции использована известная модель, учитывающая симметрию частотных спектров фотолюминесценции и поглощения [1,2] с внесенным в эту модель уточнением автора [3 – 6], согласно которому точке зеркальной симметрии этих спектров соответствует частота излучения источника, возбуждающего фотолюминесценцию.

Кроме того, следует предварительно отметить, что в дальнейшем под поглощением мы будем понимать оптические потери, связанные не только с однофотонными резонансными процессами, но мы также должны учесть процессы многофотонные, которые, в конечном счете, и определяют процесс взаимодействия излучения со средой [3 – 6].

1. Introduction.

Recently there was a significant amount of the works devoted to the description of the phenomenon of a photoluminescence in glasses and optical fiber, activated by bismuth. Interest of researchers to the given problem is caused by that circumstance that the near infra-red area of a spectrum is of special interest for the experts occupied with reception of laser generation in the field of 1.0 – 1.5 microns, and also – actual for the given area of a spectrum of possibility of transfer on optical fiber of considerable volumes of the information.

In the offered review of some the experimental results showing this phenomenon is executed, and from uniform positions representations about the process responsible for a photoluminescence in case of use of various laser sources for excitation of environment, activated by bismuth are considered.

For interpretation of process of a photoluminescence the known model considering symmetry of frequency spectra of a photoluminescence and frequency spectra of a absorption [1, 2] with specification brought in this model of the author [3 – 6] is used.

Besides, it is necessary preliminary to notice, that further we will understand the optical losses connected not only with one-photon resonant processes as absorption, but we also should consider multiphotons processes which, finally, and define interaction of radiation with the environment [3 – 6].

2. Ожидаемое отличие результатов оптических экспериментов по наблюдению фотолюминесценции, выполненных с использованием стекловолокна, от результатов эксперимента, выполненного по традиционной схеме с использованием плоских стеклянных образцов.

.
2. Expected difference of results of optical experiments on supervision of the photoluminescence, executed with optical fiber activated by bismuth use, from results of the experiment executed under the traditional scheme with use of flat glass samples activated by bismuth.

Возможность регистрации оптических эффектов, с использованием «длинных» волоконных световодов и лазерных источников излучения за счет увеличения длительности, длины и, практически, неограниченного «удельного» объема взаимодействия светового излучения со средой, обеспечивающего накопление сколь-угодно слабых процессов преобразования энергии световых квантов, обычно недоступны для наблюдения тонких эффектов в стандартных условиях, например, в случае использования в качестве образцов стеклянных пластинок.

Возможность получения с использованием протяженных световодов уникальных экспериментальных результатов по взаимодействию излучения с рассматриваемой средой, возбуждающего фотолюминесценцию последней, приводит к простым численным оценкам и наглядной интерпретации физических процессов. А, именно, удается установить связь между опирающимися на закон сохранения энергии численными оценками длины волны (частоты) максимального пика излучения фотолюминесценции и ответственными за наблюдаемые максимумы фотолюминесценции, широкоизвестными табличными величинами длин волн линий, соответствующих межуровневым переходам атомарного висмута, полученными в совершенно других экспериментальных условиях [7,8].

С другой стороны, высокая эффективность фотолюминесценции стекловолокна, активированного висмутом, позволяет сопоставлять и сравнивать результаты экспериментов, выполненных различными группами исследователей в случае использования ими одинаковых или различных источников возбуждения фотолюминесценции на одном или различных участках спектра. Такой анализ способствует выявлению и более достоверной интерпретации механизмов, ответственных за процесс фотолюминесценции активированного висмутом стекла и стекловолокна.

The possibility of registration in "long" optical fiber activated by bismuth optical effects with use laser sources of radiation at the expense of increase in duration, length and, practically, "unlimited specific" volume of interaction of light radiation with the environment, providing accumulation of weak processes of transformation of energy of light quanta, are usually inaccessible to supervision of thin effects in standard conditions, for example, in case of use as samples of glass plates activated by bismuth.

The use of extended optical fiber activated by bismuth of unique experimental results on interaction of radiation with the considered environment, raising a photoluminescence of last, leads to simple numerical estimations and evident interpretation of physical processes.

And, it is possible to establish connection between leaning against the law of conservation of energy by numerical estimations of frequency of the maximum peak of radiation of a photoluminescence and responsible for observable maxima of a photoluminescence, accessible table values of lengths of waves (Bohr jump-frequencies) of the lines corresponding to transitions of atomic bismuth, received in perfect other experimental conditions [7, 8].

On the other hand, high efficiency of a photoluminescence of the optical fiber, activated by bismuth, allows to compare results of the experiments, executed by various groups of researchers in case of use by them of identical or various sources of excitation of a photoluminescence on one or various segments of a spectrum. Such analysis promotes revealing and more authentic interpretation of the mechanisms responsible for process of a photoluminescence of glass, activated by bismuth, and optical fiber, activated by bismuth.

Целью работы является установление с учетом правила зеркальной симметрии соответствия между величинами вычисленных согласно закону сохранения энергии на основании опубликованных разными авторами экспериментальных результатов частот осцилляторов (оптических электронов атомов висмута, ответственных за фотолюминесценцию висмутсодержащей среды), с табличными значениями [7, 8] длин волн (частот переходов), характерных для спектров испускания (и, соответственно, поглощении) атомарного висмута.

Обнаруженное в результате проведенных расчетов такое совпадение величин частот (длин волн) позволяет дать ответ о природе фотолюминесцентного процесса, возбуждаемого в активированных висмутом волоконных световодах, а также в стекле. Поскольку данные об особенностях спектров фотолюминесценции извлекались из опубликованных работ с помощью угольника и линейки, что вносило ту или иную величину погрешности в используемые для расчета численные величины, не приходилось ожидать стопроцентного совпадения рассчитанных и табличных значений. Поэтому допускалось, что могло иметь место несовпадение рассчитанных и табличных значений в пределах нескольких (от 3 до 5) процентов.

В дальнейшем тексте статьи понятие «частота осциллятора» соответствует понятию боровской частоты атомарного перехода, определяемой, согласно [9], из соотношения:

The work purpose is the confirmation a rule of mirror symmetry between calculated jump-frequencies of bismuth atoms optical electrons, responsible for a photoluminescence ( the experimental results of different authors), with tabular values of [7, 8] lengths of waves (jump-frequencies of transitions), characteristic for emission spectra (and, accordingly, absorption) atomic bismuth.
Such coincidence of quantities of frequencies found out as a result of spent calculations (lengths of waves) allows to answer about the nature of the photoluminescent process, raised in fiber optical and also in glass, activated by bismuth.

As the data about features of spectra of a photoluminescence was taken from the published works by means of a square and a ruler that brought this or that size of an error in numerical sizes used for calculation, it was not necessary to expect absolute coincidence of the calculated and tabular values. Therefore it was supposed, that discrepancy of the calculated and tabular values within several ( to ~10 %) could take place.
In the further text of article concept «frequency of transition» corresponds to concept of Bohr jump-frequency of the atomic transition defined, according to [9], from a parity:

$$\nu_{ij} = (E_i - E_j)/h, \qquad (1)$$

где $\nu_{ij}$ - боровская частота межуровневого перехода атомарного висмута, связывающего (в висмуте) два электронных уровня противоположной четности;

$h$ - постоянная Планка;

$E_j$ – энергия ниже расположенного уровня атомарного висмута;

$E_i$ – энергия выше расположенного уровня атомарного висмута.

Поскольку атомы висмута внедрены в однородную аморфную изотропную среду (кварцевое стекло), температура которой существенным образом зависит от условий возбуждения фотолюминесценции, то в дальнейшем изложении не предполагается, что в атоме висмута отсчет частот переходов, используемых для возбуждения фотолюминесценции, имеет место только от невозбужденного основного уровня атомарного висмута $^4S^0_{3/2}$.

Поэтому в соответствии с распределением Больцмана можно считать, что в исследуемой среде, фактически одновременно, сосуществует множество способных к участию в процессе фотолюминесценции двухуровневых систем, определяемых боровскими частотами атомарного висмута $v_{ij} = (E_i - E_j)/h$, где $i,j = 6,7,8....$

И поэтому в случае выполнения условия резонанса (смотри ниже соотношение (3)) излучение накачки, вызывающее фотолюминесценцию в среде, содержащей атомы висмута, способно взаимодействовать с любым из двухуровневых разрешенных переходов и, соответственно, с оптическими электронами висмута, участвующими в процессе.

Совместное рассмотрение такого взаимодействия с двухуровневой средой с одновременным использованием модели классического осциллятора Лоренца, согласно принципу дополнительности Нильса Бора [10], позволяет определить собственные частоты колебаний такого осциллятора, определяемые боровскими частотами $v_{ij}$ (смотри соотношение (1)), что, в конечном счете, дополняет и расширяет представление о происходящих в среде процессах.

where $v_{ij}$ – Bohr jump - frequency of transitions of the atomic bismuth connecting two electronic levels of opposite parity;

$h$ - Planck's constant;

$E_j$ – energy below the located level of atomic bismuth;

$E_i$ – energy above the located level of atomic bismuth.

Atoms of bismuth are introduced in the homogeneous amorphous isotropic environment (quartz glass), which temperature essentially depends on conditions of excitation of a photoluminescence. It is not supposed that in atom of bismuth frequencies of the transitions used for excitation of a photoluminescence, takes place only from basic ( ground) level of atomic bismuth $^4S^0_{3/2}$. Therefore according to distribution of Boltzman it is possible to consider that in the investigated environment, actually simultaneously, the set capable to participation in process of a photoluminescence of the two-level systems defined by Bohr frequencies of atomic bismuth $v_{ij} = (E_i - E_j)/h$, where $i = 5,6,7,8…$, $j = 4,5,6,7…$ And consequently in case of performance of a condition of a resonance (look low a parity (3)) the radiation, causing a photoluminescence in the environment, containing atoms of bismuth, is capable to operate with any resolved of the two-level transitions and, accordingly, with optical electrons of the bismuth, participating in process.

Joint consideration of such interaction with the two-level environment with simultaneous use of model classical oscillator of Lorenz, according to Bohr correspondence principle [10], allows to define own frequencies of such oscillator, defined by Bohr frequencies $v_{ij}$.

Предварительно остановимся на двух особенностях процесса взаимодействия излучения, возбуждающего фотолюминесценцию в стеклянной матрице, допированной атомами висмута.

1. Первая особенность связана со значительным уширением линий взвешенного в среде вещества (атомарного висмута), и их взаимным наложением друг на друга. Представление об уширении линий висмута и их взаимном наложении можно получить анализируя спектр поглощения (оптических потерь) допированного висмутом волновода, представленный в работе [11]. Поглощение охватывает участок спектра, занимаемый атомарными линиями висмута 905,86 нм; 934,255 нм; 965,72 нм; 982,88 нм; 1010,61 нм, сливающимися в единый контур. Попутно отметим, что из указанной группы линий согласно [7] линия 965,72 нм является наиболее сильной (в атоме висмута она соответствует переходу 7s -7p ).

При обработке опубликованных различными исследовательскими группами результатов следует иметь в виду, что само положение линий (поглощения) взвешенного в кварцевом стекле или в световоде атомарного висмута, например, из-за допплеровского уширения, а также из-за смещения атомных уровней Bi вследствии воздействия на атомы висмута возбуждающего фотолюминесценцию лазерного излучения, изменяющего распределение электронов по атомным уровням [12], – нестабильно, что может привести к недостаточно точному совпадению полученных в экспериментах результатов из ниже представленного перечня работ с табличными значениями [7, 8].

В спектре поглощения (оптических потерь) [11] кварцевого волоконного световода, активированного висмутом, заслуживают внимания потери светового излучения в области 1390 нм, смещенной в высокочастотную область спектра относительно линии 1433,15 нм [7], полуширина которой превышает 100 нм. Из-за сильного «непропускания» излучения в области спектра вблизи 1390 нм наблюдение фотолюминесценции на этом участке спектра затруднено[13].

Let's preliminary stop on two features of process of interaction of the radiation raising a photoluminescence in a glass matrix, допированной atoms of bismuth.

1. The first feature is connected with considerable уширением lines weighed in the environment of substance (atomic bismuth), and their mutual imposing against each other. Representation about уширении lines of bismuth and their mutual imposing can be received analyzing an absorption spectrum (optical losses) допированного the wave guide bismuth, presented in work [11]. Absorption covers the site of a spectrum occupied with atomic lines of bismuth of 905,86 nanometers; 934,255 nanometers; 965,72 nanometers; 982,88 nanometers; 1010,61 nanometers, merging in a uniform contour. We will in passing notice that from the specified group of lines according to [7] line of 965,72 nanometers is the strongest (in atom of bismuth it corresponds to transition 7s-7p).

At processing of the results published by various research groups it is necessary to mean that position of lines (absorption) weighed in quartz glass or in an optical path of atomic bismuth, for example, because of допплеровского уширения, and also because of displacement of nuclear levels Bi вследствии influences on atoms of bismuth of the laser radiation raising a photoluminescence changing distribution электронов on nuclear levels [12], – is astable that can lead to insufficiently exact coincidence of the results received in experiments from more low presented list of works with tabular values [7, 8].

In an absorption spectrum (optical losses) [the 11] quartz fiber optical paths, activated by bismuth, are worthy losses of light radiation in the field of 1390 nanometers, displaced in high-frequency area of a spectrum concerning a line of 1433,15 nanometers [7] which semiwidth exceeds 100 nanometers. Because of strong «непропускания» radiations in the field of a spectrum near to 1390 nanometers photoluminescence supervision on this site of a spectrum is complicated [13].

. Если центральная частота линии 1433,15 нм соответствует частоте классического гармонического осциллятора Лоренца, то, согласно принципу дополнительности Н. Бора, асимметрия, связанная с дисперсией показателя преломления атомов висмута именно в этой области спектра: n < 1 или n > 1 (смотри [14,15]), способна объяснить большую величину оптических потерь с высокочастотной стороны центральной частоты; максимум потерь соответствует длине волны ~ 1390 нм – участку спектра, где там где n < 1.

2. Что касается второй особенности процесса взаимодействия излучения с исследуемой средой, то так как часть фотонов излучения накачки $v_p$ при распространении вдоль кварцевого световода, допированного атомами или наночастицами висмута, затрачивается на изменение распределения электронов атомарного висмута по уровням, что приводит к частичному изменению показателя преломления исследуемой среды, т. е. имеет место динамическая компенсация [4] показателя преломления $n_\sigma(v)$ – просветление среды на частоте накачки.

Присутствие включений атомов (или наночастиц) висмута в кварцевом стекле фактически соответствует ансамблю разночастотных "классических" осцилляторов Лоренца [14], взвешенных в стекле. Такие разнесенные по частоте осцилляторы, являются причиной того, что конкретная величина показателя преломления $n_\sigma(v_p)$ кварцевого стекла, допированного висмутом, зависит от концентрации атомов висмута $N$, напряженности поля $E$ светового излучения накачки и частотной расстройкой $\delta v$ между частотой возбуждающего излучения $v_p$ и собственной частотой колебаний $v_{ij}$ оптических электронов атомов висмута, соответствующей боровской частоте. В общем случае, для показателя преломления $n_\sigma(v_p)$ кварцевого допированного висмутом волоконного световода можно записать:

If the central frequency of a line of 1433,15 nanometers corresponds to frequency classical harmonious осциллятора Lorentz, that, according to a principle дополнительности to N.Bora, the asymmetry connected with a dispersion of an indicator of refraction of atoms of bismuth in this area of a spectrum: n <1 or n> 1 (look [14,15]), it is capable to explain the big size of optical losses from the high-frequency party of the central frequency; the maximum of losses corresponds to length of a wave ~ 1390 nanometers – to a spectrum site, where there where n <1.

2. As to the second feature of process of interaction of radiation with the investigated environment as the part of photons of radiation of a rating ν p at distribution along a quartz optical path, допированного atoms or наночастицами bismuth, is spent for distribution change электронов atomic bismuth on levels that leads to partial change of an indicator of refraction of the investigated environment, i.e. Dynamic indemnification [4] indicators of refraction n σ (ν) – an enlightenment of environment on frequency of a rating takes place.

Presence of inclusions of atoms (or наночастиц) bismuth in quartz glass actually corresponds to ensemble разночастотных "classical" осцилляторов Lorentz [14], weighed in glass. Such carried on frequency осцилляторы, are at the bottom of that the concrete size of an indicator of refraction n σ (vp) quartz glass, допированного bismuth, depends on concentration of atoms of bismuth N, intensity of field E of light radiation of a rating and frequency расстройкой δ ν between frequency of exciting radiation vp and own frequency of fluctuations ν ij optical электронов atoms of the bismuth, corresponding Bohr frequency. Generally, for a refraction indicator n σ (vp) quartz допированного bismuth of a fiber optical path it is possible to write down:

$$n_\sigma(v_p) = n(f(N,E,\delta v)) + n_{кв}., \qquad (2)$$

где $n_{кв}$ показатель преломления чистого кварцевого стекла.

Where n кв an indicator of refraction of pure quartz glass.

3.
<u>Роль закона сохранения энергии в модели, используемой для описания особенностей спектра фотолюминесценции активированной висмутом кварцевой матрицы. Фотолюминесценция и вынужденное излучение среды, рассматриваемой в рамках двухуровневой модели. Принцип дополнительности Нильса Бора и классическая модель осциллятора Лоренца.</u>

<u>Role of the law of conservation of energy in the model used for the description of features of a spectrum of a photoluminescence of a quartz matrix activated by bismuth. A photoluminescence and the compelled radiation of the environment considered within the limits of two-level model. A principle дополнительности Nilsa of the Pine forest and classical model осциллятора Lorentz.</u>

Анализируя полученные и опубликованные разными авторами спектры поглощения (оптических потерь, пропускания) световода или кварцевого стекла, активированного висмутом, следует отметить что, как и предполагалось во 2 разделе настоящей статьи, эти спектры, например, за счет эффекта накопления, заметно отличаются друг от друга (сравни, например, результаты, опубликованные в [11,16]).

При изучении оптических потерь в волоконном световоде необходимо обращать внимание на такие участки спектра, на которых в элементарном акте взаимодействия излучения с двухуровневой средой, возможно участие несколько фотонов [4]. Переход атомов висмута из нижнего («невозбужденного») состояния в верхнее («возбужденное») состояние, согласно закону сохранения энергии, происходит в соответствии с основным соотношением фотолюминесценции, подробно рассмотренным автором в работах [3-6]:

$$h \nu_i = 2 h \nu_p - h \nu_{ij} \qquad (3)$$

Такой переход атома на более высокий «возбужденный» уровень связан с изменением энергии поглощающего атома на величину, определяемую соотношением $\delta E = h \nu_{ij}$.

В соотношении (3):

$h \nu_p$ - энергия фотонов возбуждающего излучения;

$\nu_p$ - частота фотонов возбуждающего излучения;

$\nu_i$ - частота фотонов фотолюминесценции;

$\nu_{ij}$ – боровская «резонансная» частота межуровневого перехода атомарного висмута, помещенного в однородную изотропную среду (кварцевое стекло), связывающего нижний (невозбужденный, но необязательно основной) и возбужденный уровни атома висмута;

$h$ -постоянная Планка.

Соотношение (3) может быть преобразовано к виду:

Analyzing the spectra of absorption received and published by different authors (optical losses, пропускания) an optical path or the quartz glass activated by bismuth, it is necessary to note that, as well as it was supposed in 2 section of present article, these spectra, for example, at the expense of effect of accumulation, considerably differ from each other (compare, for example, the results published in [11,16]).

At studying of optical losses in a fiber optical path it is necessary to pay attention to such sites of a spectrum on which in the elementary certificate of interaction of radiation with the two-level environment, participation some photons [4] is possible. Transition of atoms of bismuth from the bottom ("not raised") condition in the top ("raised") condition, according to the law of conservation of energy, occurs according to the basic parity of the photoluminescence, in detail considered author in works [3-6]:

h ν i = 2 h νp - h ν ij (3)

Such transition of atom to higher "raised" level is connected with change of energy of absorbing atom on the size defined by a parity δE = h ν i j.

In the ratio (3):

h νp - energy of photons of exciting radiation;

νp - frequency of photons of exciting radiation;

ν i - frequency of photons of a photoluminescence;

ν i j – Bohr "resonant" frequency межуровневого transition of the atomic bismuth placed in the homogeneous isotropic environment (quartz glass), connecting bottom (not raised, but it it is unessential the basic) and raised levels of atom of bismuth;

h - Planck's constant.

The parity (3) can be transformed to a kind:

$$h \nu_{ij} - h \nu_p = h \nu_p - h \nu_i . \qquad (3a)$$

В этом случае становится наглядным тезис о симметрии спектров фотолюминесценции и поглощения исследуемой среды.

Специально отметим, что соотношение (3) устанавливает линейную зависимость между числом фотонов фотолюминесценции и числом фотонов возбуждающего излучения. В работах, посвященных исследованию зависимости величины люминесцентного сигнала от величины интенсивности возбуждающего излучения, такая линейная зависимость, как правило, отмечается.

Для справки приведем соотношение, представляющее закон сохранения энергии для элементарного акта вынужденного излучения в случае инвертированной среды,

$$h\nu_{ij} + h\nu^*_{ij} = 2h\nu_{ij}, \quad (4)$$

согласно которому в элементарном акте при взаимодействии с возбужденным атомом удвоение $2h\nu_{ij}$ числа квантов светового излучения происходит по линейному закону и имеет место линейная зависимость между числом квантов «возбуждающего» излучения $h\nu_{ij}$ и числом квантов покидающего среду излучения за счет запасенной средой инверсии (обозначение величины $E^* = h\nu^*_{ij}$ указывает на то, что переход $ij$ инвертирован). В настоящей работе такие процессы не рассматриваются. Согласно [4], необходимо четко различать между собой процесс вынужденного излучения, описываемый соотношением (4), и процесс фотолюминесценции, который соответствует соотношению (3).

In this case there is evident a thesis about symmetry of spectra of a photoluminescence and absorption of the investigated environment.

Let's specially notice that the parity (3) establishes linear dependence between number of photons of a photoluminescence and number of photons of exciting radiation. In the works devoted to research of dependence of size of a luminescent signal from size of intensity of exciting radiation, such linear dependence, as a rule, is marked.

For the inquiry we will result the parity representing the law of conservation of energy for the elementary certificate of compelled radiation in case of the inverted environment,

h ν ij + h ν * ij = 2 h ν ij, (4)

According to which in the elementary certificate at interaction doubling 2 h ν ij numbers of quanta of light radiation occurs to the raised atom under the linear law and linear dependence between number of quanta of "exciting" radiation h ν ij and number of quanta of radiation leaving the environment at the expense of the inversion reserved by the environment (the designation of size E* =h ν * ij specifies that transition ij is inverted) takes place. In the present work such processes are not considered. According to [4], it is necessary to distinguish accurately among themselves the process of the compelled radiation described by a parity (4), and process of a photoluminescence which corresponds to a parity (3).

4. <u>Анализ опубликованных в научной периодике экспериментальных результатов по фотолюминесценции волоконных световодов, активированных висмутом и возбуждаемых с помощью источников лазерного излучения с различными длинами волн.</u>

The analysis of the experimental results published in the scientific periodical press on a photoluminescence of the fiber optical paths activated by bismuth and raised by means of sources of laser radiation with various lengths of waves.

Предлагаемая в настоящей работе интерпретация явления фотолюминесценции основывается на давно устоявшемся и сложившемся в пятидесятые годы двадцатого века представлении, связанном с характерным явлением зеркальной симметрии между спектром фотолюминесценции и спектром поглощения исследуемой среды, известном как закон Левшина В. Л. - Ломмеля. [1,2]. Эта закономерность, как правило, отмечается при наблюдении и исследовании фотолюминесценции красителей или иных молекулярных, а также наноразмерных сред. В работах [3,5,6] для случая наноразмерного кремния представлено уточненное автором детальное описание физической модели фотолюминесценции, учитывающее указанную симметрию. Основное уточнение связано с тем обстоятельством, что при построении графической зависимости фотолюминесценции от частоты (энергии фотонов) светового излучения, точка, относительно которой рассматривается зеркальная симметрия спектров, определяется частотой $\nu_p$ возбуждающего фотолюминесценцию излучения. На этом этапе и появляются фотоны (кванты) фотолюминесценции, частота которых подчиняются соотношению (3).

Именно относительно частоты возбуждающего излучения $v_p$ и ведётся поиск симметричных частот: частоты фотолюминесцентного излучения $v_i$, и боровской частоты $v_{ij}$, характеризующей поглощение среды (или оптические потери).

Следует отметить, что указанная симметрия не всегда может быть установлена и обнаружена в явном виде из-за неодинаковой чувствительности регистрирующих фотоприемников, используемых для получения спектров поглощения и спектров фотолюминесценции, находящихся в разных участках спектра, или вследствии того, что часть излучения фотолюминесценции может быть поглощена средой, что так же может внести искажения в симметрию.

Однако, в общем случае, в кварцевой матрице, допированной висмутом, структура спектра поглощения может оказаться достаточно сложной, так как происходит одновременное взаимодействие возбуждающего среду излучения на частоте $v_p$ со значительным количеством отдельно стоящих уширенных линий висмута $v_{ij}$ и поэтому, как спектр поглощения, так и спектр фотолюминесценции оказываются уширенными на значительную величину. Использование разными группами исследователей для возбуждения фотолюминесценции лазерных источников в различных областях спектра, а также присутствие значительного числа переходов в атомарном висмуте открывает возможность появления излучения фотолюминесценции в широком спектральном интервале, на отдельных участках которого из-за поглощения атомарным висмутом наблюдение фотолюминесценции может быть затруднено или даже нереализуемо.

] Interpretation of the phenomenon of a photoluminescence offered in the present work is based on the representation which has for a long time settled and developed in the fifties the twentieth century connected with the characteristic phenomenon of mirror symmetry between a spectrum of a photoluminescence and a spectrum of absorption of the environment investigated, known as Levshin V. L's law - Lommelja. [1,2]. This law, as a rule, is marked at supervision and research of a photoluminescence of dyes or others molecular, and also наноразмерных environments. In works [3,5,6] for a case наноразмерного silicon the detailed description of physical model of the photoluminescence specified by the author, considering the specified symmetry is presented. The basic specification is connected with that circumstance that at construction of graphic dependence of a photoluminescence from frequency (energy of photons) light radiation, the point concerning which mirror symmetry of spectra is considered, is defined by frequency vp radiation raising a photoluminescence. At this stage also there are photons (quanta) of the photoluminescence which frequency submit to a parity (3).

Concerning frequency of exciting radiation vp search of symmetric frequencies also is conducted: frequencies of photoluminescent radiation v i, and Bohr frequency v ij, environment characterising absorption (or optical losses).

It is necessary to notice that the specified symmetry not always can be established and found out in an explicit form because of unequal sensitivity of the registering photodetectors used for reception of spectra of absorption and spectra of a photoluminescence, being in different sites of a spectrum, or вследствии that the part of radiation of a photoluminescence can be absorbed environment that as can bring distortions in symmetry.

However, generally, in a quartz matrix, допированной bismuth, the structure of a spectrum of absorption can appear enough difficult as there is a simultaneous interaction of radiation raising the environment on frequency vp to a significant amount of the separate widened lines of bismuth v ij and consequently, both the absorption spectrum, and a photoluminescence spectrum appear widened on considerable size. Use by different groups of researchers for excitation of a photoluminescence of laser sources in various areas of a spectrum, and also presence of considerable number of transitions at atomic bismuth opens possibility of occurrence of radiation of a photoluminescence in a wide spectral interval on which separate sites because of absorption by atomic bismuth photoluminescence supervision can be complicated or even нереализуемо.

Далее со ссылками на опубликованные работы следует **таблица**, в которой в соответствующих столбцах представлены следующие величины:

$v_i$ - частота, соответствующая максимально высокому пику кривой фотолюминесценции;

$v_p$ - частота возбуждающего излучения;

$v_{ij}$ - боровская частота электронного перехода, ответственного за наблюдаемое в эксперименте максимальное значение (пик) фотолюминесценции атомарного висмута, помещенного в однородную изотропную среду (кварцевое стекло);

$\lambda_{ij}$ - оценочная велична длины волны, характеризующей указанный переход, ответственный за максимум кривой фотолюминесенции;

$\lambda_t$ - табличное значение длины волны, определяющей электронный переход в атоме висмута, отвественный за фотолюминесценцию для выбранной длины волны возбуждающего излучения;

$h$ - постоянная Планка.

Точность совпадений величин $\lambda_{ij}$ и $\lambda_t$ связана с вызывающими сдвиг и уширение линий процессами (допплер, столкновительные, деформационные, температурные), поэтому в таблице, как правило, имеет место их несовпадение в третьем или четвертом знаке.

Еще одной причиной отсутствия точного совпадения указанных величин может быть сам процесс извлечения численных значений максимальных значений спектральных кривых фотолюминесценции из опубликованных в литературе графических зависимостей и результатов.

Further with references to the published works the table in which in corresponding columns following sizes are presented follows:

v i - the frequency corresponding to as much as possible high peak of a curve photoluminescence;

vp - frequency of exciting radiation;

v ij - Bohr frequency of the electronic transition responsible for maximum value (peak) observed in experiment of a photoluminescence of atomic bismuth, placed in the homogeneous isotropic environment (quartz glass);

λ ij - estimated велична lengths of the wave characterising specified transition, responsible for a curve maximum фотолюминесенции;

λ t - tabular value of length of the wave defining electronic transition in atom of bismuth, отвественный for a photoluminescence for the chosen length of a wave of exciting radiation;

h - Planck's constant.

Accuracy of coincidence of sizes λ ij and λ t is connected with causing shift and уширение lines processes (допплер, столкновительные, deformation, temperature), therefore in the table their discrepancy in the third or fourth sign, as a rule, takes place.

Process of extraction of numerical values of the maximum values of spectral curves of a photoluminescence of the graphic dependences published in the literature and results can be one more reason of absence of exact coincidence of the specified sizes.

**Таблица .**

<u>Соотношение между табличным значением длины волны $\lambda_t$ линии атомарного висмута и ее рассчитанной величиной на основании учета положения максимального пика фотолюминесценции волоконного световода, активированного висмутом, для используемого источника возбуждения фотолюминесценции.</u>

<u>Parity between tabular value of length of a wave λ t lines of atomic bismuth and its calculated size on the basis of the account of position of the maximum peak of a photoluminescence of the fiber optical path activated by bismuth, for a used source of excitation of a photoluminescence.</u>

Reference number
In the list
Literatures. Max. Peak fotoljum-tsii (length of a wave, frequency λi, ν i) - an estimation according to an illustration presented by authors tsiti-ruemoj of work. Length of a wave of a laser source of radiation of excitation
 λ p; (frequency - ν p). Calculated
On the basis of published эксперим.результатов
Length of a wave
 λ ij and frequency of transition - ν ij, responsible for макс. Peak fotoljum-tsii. Tabular [7] size of length of a wave λ t, (in line brackets-brightness it agree [7]).

| Номер ссылки в списке литературы. | Макс. пика фотолюм-ции (длина волны, частота $\lambda_i$, $\nu_i$) - оценка согласно иллюстрации, представленной авторами цити-руемой работы. | Длина волны лазерного источника излучения возбуждения $\lambda_p$ ; (частота - $\nu_p$). | Рассчитанная на основании опубликованных эксперим.результатов длина волны $\lambda_{ij}$ и частота перехода - $\nu_{ij}$, ответственного за макс. пика фотолюм-ции. | Табличная [7] величина длины волны $\lambda_t$, (в скобках-яркость линии согласно [7]). |
|---|---|---|---|---|
| [11] | ~ 1132 нм; 8834 см$^{-1}$ | 1075 нм; 9302,33 см$^{-1}$ | 1024 нм, 9770 см$^{-1}$ | 965,72нм, (2000) |
| [17] | ~757 нм; 13210 см$^{-1}$ | 532 нм; 18797 см$^{-1}$ | 410,0 нм, 24385 см$^{-1}$ | 412,184 нм, (14), 412,152нм, (14) |
| [17] | 1164 нм; 8588,9 см$^{-1}$ | 532 нм; | 340,5 нм 29365 см$^{-1}$ | 340,52нм,(60); |
| [17] | ~1164 нм 8588,9 см$^{-1}$ | 925 нм; 10810 см$^{-1}$ | 767,3 нм 13032 см$^{-1}$ | 783,6 нм,(400); 750,2 нм,(6); |
| [17] | ~1600 нм 6250 см$^{-1}$ | 925 нм | 650,5 нм 15371,6 см$^{-1}$ | 613,4нм, (50) |
| [17] | ~ 1390 нм 7194,2 см$^{-1}$ | 800 нм 12500 см$^{-1}$ | 561,6 нм 17806 см$^{-1}$ | 555,2 нм(1) 574,2нм(30) |
| [17] | ~ 1351 нм 7402 см$^{-1}$ | 1230 нм; 6130 см$^{-1}$ | 1128, 9 нм 8858 см$^{-1}$ | 1107,3нм(15) |
| [17] | ~ 1425 нм 7017 см$^{-1}$ | 1230 нм | 1082нм 9242 см$^{-1}$ | 1054,0нм (8) |
| [13] | 1693,5 нм 5904 см$^{-1}$ | 1230 нм 8130 см$^{-1}$ | | 965,7нм(2000) 10355,1 см$^{-1}$ |
| [13] | 1643,17нм 6085,8 см$^{-1}$ | 1230 нм | | 982,88 нм(300) 10174,2 см$^{-1}$ |
| [13] | 1571,1 нм 6365 см$^{-1}$ | 1230 нм | | 1010,6 нм(20) 9895 см$^{-1}$ |
| [13] | 1130 нм 8849,55 см$^{-1}$ | 1058 нм 9451,79 см$^{-1}$ | 994,6 нм 10054 см$^{-1}$ | 982,8 нм(300) |
| [18,19] | 1100 нм; 9090 см$^{-1}$ | 514 нм; | 335,3 нм 10054 см$^{-1}$ | 339,72 нм (55) 323,9 нм(10) |
| [18,19] | 720 нм | 514 нм | 399,6 нм | 388,8 нм (40) |
| [20,21] | 750 нм | 500 нм | 374,99нм | 359,6нм(38); 388,8 нм(40) |
| [20,21] | 1140нм; 8772 см$^{-1}$ | 500 нм; | 320,2нм | 314,4нм(8); 323,9 нм(10) |
| | 1122нм; 8913 см$^{-1}$ | 700 нм; | 508,7нм | 472,2 нм(60) |
| | 1250 нм; 8000 с м$^{-1}$ | 800 нм; | 588,2нм | 574,2нм(30) |
| [20,22] | 1300 нм; 7692 см$^{-1}$ | 800 нм; | 577,7нм | 574,2нм (30) |
| [20,23] | 1315 нм; 7604,5 см$^{-1}$ | 808 нм; | 583,1нм | 574,2 нм (30) |
| [20,24] | 1310 нм; 7634 см$^{-1}$ | 808 нм; | 584,1нм | 574,2 нм(30) |
| [20,25] | 1150 нм 8696 см$^{-1}$ | 980 нм; | 853,8нм | 854,4 нм(40); |
| [20,26] | 1210 нм; 8264 см$^{-1}$ | 405 нм | 243,2 нм | 243,3нм(30); |
| | 1173 нм; 8525 см$^{-1}$ | 514 нм | 329,1 нм | 324,0нм(10); |
| | 1300нм; 7692 см$^{-1}$ | 808 нм | 586,1 нм | 574,2нм(30) |
| [20,27] | 1100 нм; | 700 нм | 513,3 нм | 472,2нм(60) |

|  | 9091 см$^{-1}$ |  |  |  |
|  | 1250 нм ;<br>8000 см$^{-1}$ | 800 нм | 588,2 нм | 574,2нм(30) |

Ниже – результаты, полученные для случая фотолюминесценции, наблюдаемой в стеклянной плоской пластине, допированной висмутом. Использование перестраиваемого лазерного источника для возбуждения в исследуемом образце фотолюминесценции дополняет представление о зеркальной симметрии между спектрами фотолюминесценции и поглощения.

| [16] | ~ 1153,5 нм | 502 нм | 314,9 -322,7 нм | 306,7 нм (3600) |
| [16] | ~1153,5 нм | 525 нм | 339,8 нм | 339,7 нм (55) |
| [16] | ~1085,4 нм | 680 нм | 472 нм | 472,2 нм (3×30) |
| [16] | ~1171,6 нм | 738 нм | 540 нм | 555,2 нм (1)<br>527,4 нм (40) |
| [16] | ~1260 нм | 798 нм | 584 нм | 574 ,2 нм (30) |

Прежде всего следует обратить внимание на два правых столбца таблицы.

В правом столбце таблицы согласно табличным данным, представленным в работе [7] использованы величины длин волн линий висмута, наиболее близко находящиеся к вычисленным величинам. В предыдущем столбце указаны значения длин волн, полученные с использованием соотношения (3) на основании анализа полученных разными авторами результатов по наблюдению фотолюминесценции в активированных висмутом волоконных световодах [11,13,17-26] и стеклянных пластинах [16]. Невозможностью точного определения максимума пика фотолюминесценции на основании экспериментальных кривых, представленных в упомянутых выше работах можно объяснить 3-5 % несовпадение табличных значений линий и величин длин волн линий, рассчитанных из представленных в экспериментальных работах [11,13,17-26] результатов. О причинах недостаточной точности совпадения табличных и рассчитанных величин упоминалось выше в разделах 1-3 данной работы.

Тем не менее основополагающий тезис о зеркальной симметрии пиковых значений кривых фотолюминесценции, с одной стороны, а с другой – линий атомарного неионизованного висмута, с которыми может быть связано поглощение светового излучения, согласно представленным в таблице результатам подтверждается. Заслуживает внимания то обстоятельство, что множеству наблюдаемых разными авторами пиков фотолюминесценции в "зеркальном" отражении соответствует множество боровских частот (электронных переходов в атомарном висмуте), с которыми и связано множество линий атомарного висмута [7].

Естественным оказывается наличие признаков симметрии между участками спектров поглощения (оптических потерь) и фотолюминесценции, расположенными в непосредственной близости друг к другу и расположенными симметрично относительно частоты источника возбуждающего фотолюминесценцию излучения. Для алюмосиликатного допированного висмутом световода такие спектры представлены в работе [11]; они отлично подтверждают их зеркальную симметрию. Следует выделить пик в спектре поглощения вблизи линии атомарного висмута 965,71 нм, о возможных причинах уширения которого мы упоминали во 2 разделе настоящей статьи.

First of all it is necessary to pay attention to two right columns of the table.

In the right column of the table according to the tabular data presented in work [7] sizes of lengths of waves of lines of the bismuth, most close being to the calculated sizes are used. In the previous column values of lengths of the waves, the parities received with use (3) on the basis of the analysis of the results received by different authors on photoluminescence supervision in the fiber optical paths activated by bismuth [11,13,17-26] and glass plates [16] are specified. Impossibility of exact definition of a maximum of peak of a photoluminescence on the basis of the experimental curves presented in works mentioned above it is possible to explain 3-5 % discrepancy of tabular values of lines and sizes of lengths of waves of the lines calculated from [11,13,17-26] results presented in experimental works. The reasons of insufficient accuracy of coincidence of the tabular and calculated sizes it was mentioned above in sections 1-3 of the given work.

Nevertheless the basic thesis about mirror symmetry of peak values of curves of a photoluminescence, on the one hand, and with another – lines atomic неионизованного bismuth with which absorption of light radiation can be connected, according to the results presented in the table proves to be true. That circumstance is worthy that to set of peaks of a photoluminescence observed by different authors in "mirror" reflexion there corresponds set of Bohr frequencies (electronic transitions in atomic bismuth) with which the set of lines of atomic bismuth [7] is connected.

Natural there is a presence of signs of symmetry between sites of spectra of absorption (optical losses) and the photoluminescences, located in immediate proximity to each other and symmetrized concerning frequency of a source of radiation raising a photoluminescence. For алюмосиликатного допированного such spectra are presented by optical path bismuth in work [11]; they perfectly confirm their mirror symmetry. It is necessary to allocate peak in a spectrum of absorption near to a line of atomic bismuth of 965,71 nanometers, уширения which we mentioned the possible reasons in 2 section of present article.

Зеркальное симметричное смещение пика спектра фотолюминесценции и пика спектра поглощения относительно частоты возбуждающего излучения $\nu_p$ = 9302,3 см$^{-1}$ (1075 нм) в работе [11] составляет величину, равную приблизительно +/- 665 см$^{-1}$. Сложная структура спектра поглощения, состоящего из нескольких линий, и недостаточно точное совпадение между максимальной величиной пика спектра поглощения и центром атомарной линии висмута 965,71 нм может быть обусловлено сильным неоднородным уширением линий висмута, помещенного в стеклянную матрицу волоконного световода в качестве допирующей добавки, а также возможным сдвигом [12] линий атомарного висмута в поле лазерного источника накачки.

Следует отметить, что спектры поглощения (оптических потерь, пропускания) для случаев активированного висмутом волоконного световода [11,20] и плоского стеклянного образца [16], могут отличаться друг от друга, лишь в общих чертах отражая структуру друг друга. Как было указано в разделе 2 настоящей статьи, специфика спектров пропускания волоконных световодов за счет длины и объема взаимодействия излучения с веществом позволяет получать, с одной стороны, более тонкие детали спектров, чем в случае плоского кварцевого образца. С другой стороны, эффекты накопления по длине волоконного образца могут приводить за счет многофотонного процесса (3) к расплыванию и размытию в спектре поглощения контуров и структуры линий.

Поэтому в случае плоского образца не удается в полной мере получить и зафиксировать множество линий фотолюминесценции, которые наблюдаются в случае волоконного световода.

Соотношение (3) иллюстрирует то обстоятельство, что часть фотонов излучения накачки $\nu_p$ при распространении вдоль световода расходуется на обеспечение динамической компенсации дисперсии показателя преломления $n_\sigma(\nu)$ кварцевого стекла, допированного атомами или наночастицами висмута. Такие включения в стекле фактически соответствует присутствию ансамбля разночастотных "классических" осцилляторов Лоренца [14], взвешенных в стекле.

Mirror symmetric displacement of peak of a spectrum of a photoluminescence and peak of a spectrum of absorption concerning frequency of exciting radiation ν p = 9302,3 sm-1 (1075 nanometers) in work [11] makes the size equal approximately + / - 665 sm-1. The difficult structure of a spectrum of the absorption consisting of several lines, and insufficiently exact coincidence between the maximum size of peak of a spectrum of absorption and the centre of an atomic line of bismuth of 965,71 nanometers can be caused strong non-uniform уширением lines of the bismuth placed in a glass matrix of a fiber optical path in quality допирующей of the additive, and also possible shift of [12] lines of atomic bismuth in the field of a laser source of a rating.

It is necessary to notice that absorption spectra (optical losses, пропускания) for cases of the fiber optical path activated by bismuth [11,20] and the flat glass sample [16], can differ from each other, only in general reflecting structure each other. As it has been specified in section 2 of present article, specificity of spectra пропускания fiber optical paths at the expense of length and volume of interaction of radiation with substance allows to receive, on the one hand, more thin details of spectra, than in case of the flat quartz sample. On the other hand, effects of accumulation on length of the fiber sample can result at the expense of multiphoton process (3) to расплыванию and to degradation in a spectrum of absorption of contours and structures of lines.

Therefore in case of the flat sample it is not possible to receive and fix to the full set of lines of a photoluminescence which are observed in case of a fiber optical path.

The parity (3) illustrates that circumstance that the part of photons of radiation of a rating ν p at distribution along an optical path is spent for maintenance of dynamic indemnification of a dispersion of an indicator of refraction n σ (ν) quartz glass, допированного by atoms or наночастицами bismuth. Such inclusions in glass actually corresponds to ensemble presence разночастотных "classical" осцилляторов Lorentz [14], weighed in glass.

Наличие таких разнесенных по частоте осцилляторов, взвешенных в стекле, приводит к тому, что величина показателя преломления (2) кварцевого стекла, допированного висмутом, при перестройке частоты возбуждающего излучения $\nu_p$ постоянно изменяется и зависит от концентрации атомов висмута, интенсивности излучения накачки и частотной расстройкой между частотой возбуждающего излучения $\nu_p$ и собственной частотой колебаний $\nu_{ij}$, оптических электронов атомов висмута.

Фотолюминесценция сопровождает динамическую компенсацию показателя преломления среды с помещенными в неё осцилляторами (см соотношение (4)) и сама, фактически, свидетельствует о такой компенсации и просветлении среды.

Собственная частота колебаний оптических электронов во взвешенных в кварцевой матрице атомах (мезоатомах, наночастицах) висмута $\nu_{ij}$, согласно принципу дополнительности Нильса Бора [10], соответствует боровской частоте дипольного перехода между соответствующими энергетическими уровнями висмута. Напомним, что, согласно классической теории дисперсии [14,15], для таких осцилляторов (до включения излучения накачки) с низкочастотной стороны, если $\nu < \nu_{ij}$ с уменьшением величины расстройки между частотой возбуждающего излучения $\nu_p$ и боровской частотой $\nu_{ij}$ показатель преломления неограниченно возрастает $n(\nu) \gg 1$, а с высокочастотной стороны справедливо соотношение $n(\nu) < 1$.

Действительно, после включения излучения накачки благодаря процессу, описываемому соотношением (3), населенности нижнего и верхнего уровней электронного перехода атомов висмута должны выравниваться, что обеспечивает изменение показателя преломления (2) среды $n_\sigma(\nu_p) \to n_{кв}(\nu_p)$ на участке спектра, соответствующем частоте $\nu_p$ возбуждающего среду излучения. Величина $n_{кв}(\nu_p)$ соответствует показателю преломления кварцевого стекла в отсутствие допирующей добавки висмута.

"Исправленная" на данном переходе с помощью поглощения части фотонов возбуждающего излучения среда позволяет проникать на значительную глубину внутрь световода тем фотонам пучка излучения накачки, которые не участвуют в процессе (3) на входе в среду, что, однако, не препятствует дальнейшему поэтапному преобразованию

$\nu_p \to \nu_i$ при распространении оставшейся части возбуждающего излучения вдоль световода.

Presence such carried on frequency осцилляторов, weighed in glass, leads to that size of an indicator of refraction (2) quartz glasses, допированного bismuth, at reorganisation of frequency of exciting radiation ν p constantly change and depends on concentration of atoms of bismuth, intensity of radiation of a rating and frequency расстройкой between frequency of exciting radiation ν p and own frequency of fluctuations ν ij, optical электронов atoms of bismuth.

The photoluminescence accompanies dynamic indemnification of an indicator of refraction of environment with placed in it осцилляторами (sm a parity (4)) and itself, actually, testifies to such indemnification and an enlightenment of environment.

Own frequency of fluctuations optical электронов in the atoms weighed in a quartz matrix (mesoatoms, наночастицах) bismuth ν i j, according to a principle дополнительности Nilsa of the Pine forest [10], corresponds to Bohr frequency дипольного transition between corresponding power levels of bismuth. We will remind that, according to the classical theory of a dispersion [14,15], for such осцилляторов (before inclusion of radiation of a rating) from the low-frequency party if ν <ν i j with size reduction расстройки between frequency of exciting radiation ν p and Bohr frequency ν i j the refraction indicator beyond all bounds increases n (ν)>> 1, and from the high-frequency party fairly parity
n (ν) <1.

Really, after inclusion of radiation of a rating thanks to the process described by a parity (3), densities of population of the bottom and top levels of electronic transition of atoms of bismuth should be levelled that provides change of an indicator of refraction (2) environments n σ (νp) → n кв (νp) on the site of a spectrum corresponding to frequency νp of radiation raising the environment. The size n кв (νp) corresponds to an indicator of refraction of quartz glass in absence допирующей bismuth additives.

"Corrected" on the given transition by means of absorption of a part of photons of exciting radiation environment allows to get on considerable depth in an optical path to those photons of a bunch of radiation of a rating which do not participate in process (3) on an input on Wednesday that, however, does not interfere with the further stage-by-stage transformation
νp → ν i at distribution of the rest of exciting radiation along an optical path.

Появление в выходном излучении, согласно соотношению (3), фотонов фотолюминесценции на частоте $v_i$, частота которых отлична от частоты возбуждающего среду излучения $v_p$, свидетельствует об изменении энергии фотонов и возможном изменении направления распространения фотонов на частоте излучения фотолюминесценции. Как правило, в объемных образцах излучение фотолюминесценции распространяется в телесный угол 4 π стерадиан.

Истинные величины показателя преломления $n_σ (v)$, характеризующие среду с включениями атомов (наночастиц) висмута, могут быть рассчитаны на основании классической теории дисперсии, что не является предметом настоящего сообщения.

Возможность использования предлагаемого подхода опирается на принцип соответствия и условие дополнительности Нильса Бора [10], согласно которому некоторые свойства двухуровневой среды, характеризующейся определенной боровской частотой [9], могут быть смоделированы и дополнены выводами, следующими из модели классического гармонического осциллятора Лоренца [14].

Попутно отметим, что с помощью предлагаемой модели удается объяснить природу антистоксова крыла спектра фотолюминесценции (в случае, если $v_p > v_{ij}$), а также предсказать область максимального по интенсивности выхода фотолюминесценции при изменении длины волны источника, возбуждающего фотолюминесценцию, используя для оценки частоты $v_p$ источника возбуждающего фотолюминесценцию излучения согласно соотношению (3а) результаты расчета среднеарифметического значения между боровской частотой, соответствующей разрешенному в атомарном висмуте переходу, и ожидаемой или востребованной частотой пика фотолюминесценции, как правило, не совпадающей с частотой другого перехода. Такое совпадение может вызвать тушение фотолюминесценции.

Occurrence in target radiation, according to a parity (3), photoluminescence photons on frequency ν i which frequency is distinct from frequency of radiation raising the environment νp, testifies to change of energy of photons and possible change of a direction of distribution of photons on frequency of radiation of a photoluminescence. As a rule, in volume samples photoluminescence radiation extends in a space angle 4 π стерадиан.

True sizes of an indicator of refraction n σ (ν), characterising the environment with inclusions of atoms (наночастиц) bismuth, can be calculated on the basis of the classical theory of a dispersion that is not a subject of the present message.

Possibility of use of the offered approach leans against a principle of conformity and a condition дополнительности Nilsa of the Pine forest [10] according to whom some properties of the two-level environment characterised by certain Bohr frequency [9], can be simulated and added by the conclusions following from model classical harmonious осциллятора of Lorentz [14].

Let's in passing notice that by means of offered model it is possible to explain the nature антистоксова a wing of a spectrum of a photoluminescence (in case νp> ν ij), and also to predict area maximum on intensity of an exit of a photoluminescence at change of length of a wave of the source raising a photoluminescence, using for a frequency estimation νp a source of radiation raising a photoluminescence according to a parity (3a) results of calculation of arithmetic-mean value between the Bohr frequency corresponding to transition resolved in atomic bismuth, and expected or demanded frequency of peak of a photoluminescence, as a rule, not coinciding with frequency of other transition. Such coincidence can cause photoluminescence suppression.

5.Заключение.

В предлагаемом обзоре выполнена систематизация полученных и опубликованных отечественными и зарубежными исследователями некоторых экспериментальных результатов, демонстрирующих явление фотолюминесценции в стеклах и волоконных световодах, активированных висмутом, и с единых позиций рассмотрены представления о процессе, ответственном за фотолюминесценцию в случае использования для возбуждения этой среды различных лазерных источников. При интерпретации процесса фотолюминесценции использована известная модель, учитывающая симметрию частотных спектров фотолюминесценции и поглощения с внесенным в эту модель ранее опубликованным уточнением автора, согласно которому точке зеркальной симметрии этих спектров соответствует частота излучения источника, возбуждающего фотолюминесценцию.

Работа была представлена в марте 2010 года на состоявшемся в Москве и Троицке 24 съезде по спектроскопии [28].

5.Conclusion.

In the offered review ordering received and published by domestic and foreign researchers of some experimental results showing the phenomenon of a photoluminescence in glasses and fiber optical paths, activated by bismuth is executed, and from uniform positions representations about the process responsible for a photoluminescence in case of use for excitation of this environment of various laser sources are considered. At interpretation of process of a photoluminescence the known model considering symmetry of frequency spectra of a photoluminescence and absorption with earlier published specification brought in this model of the author according to which to a point of mirror symmetry of these spectra there corresponds frequency of radiation of the source raising a photoluminescence is used.

Work has been presented in March, 2010 on taken place in Moscow and Troitsk 24 congress on spectroscopy [28].

The author expresses gratitude to Plotnichenko V. G. for the publications specified to it on photoluminescence experimental researches in the wave guides activated by bismuth and to Krasovsky V. I. for support at a writing of the given work.

Библиографические ссылки.

1. А. А. Бабушкин, П. А Бажулин, Ф. А. Королев, Л. В. Левшин, В. К. Прокофьев, А. Р Стриганов,. *Методы спектрального анализа*. Под ред. В. Л. Левшина ( М.: Изд-во МГУ, 1962).

2. В. Л. Левшин, *Фотолюминесценция жидких и твердых тел* (М.Л: ГИИТЛ, 1951).

3. В. Е. Оглуздин, *Краткие сообщения по физике №12* ( М.: Фиан, стр.3, 2003).
    V. E. Ogluzdin, Bulletin of the Lebedev Physics Institute, No **12**, 1(2003).

4. В. Е. Оглуздин, *УФН* **176** 415(2006).
    V. E. Ogluzdin, Physics-Uspekhi, vol. **49**, No. 4, 401(2006).

5. В. Е. Оглуздин, *ФТП* **39** 920 (2005).
    V. E. Ogluzdin, Semiconductors, vol.**39**, 884(2005).

6. В. Е. Оглуздин, *Известия РАН, серия физическая* **70** 418 (2006).
    V. E. Ogluzdin,"Bulletin of Russian Academy of Science", Physics, Vol.**3**, 475(2006).

7. А Н Зайдель, *Таблицы спектральных линий* (М.: Наука, 1977).

8. А. С. Яценко, *Диаграммы Гартмана нейтральных атомов*. Под ред. Раутиана С Г (Новосибирск,: ВО: Наука,1993).

9. Э. Ферми, *Лекции по квантовой механике.*
    E. Fermi, Notes on Quantum Mechanics: A Course Given by Enrico Fermi at the University of Chicago ( University of Chicago Press, Chicago, 1965).

10. Н. Бор, *Атомная физика и человеческое познание* (М.: ИЛ, 1961).
    N. Bohr, Atomic Theory and the Description of Nature ( Cambridge University Press, Cambridge, 1934).

11. А. А. Крылов, П. Г. Крюков, Е. М. Дианов, О. Г. Охотников, М. Гуина Квантовая электроника, **39** (№ 1) 21 (2009).

12. V. E. Ogluzdin, Laser Physics, **16**, №8 1178-1183 (2006).

13. E.M Dianov, S.V. Firstov, O.I. Medvedkov, I. A. Bufetov, V. F. Khopin, A. N. Guryanov. Optical Fiber Communication Conference, San Diego, CA, USA, March 22- 26, 2009 (OFC2009), paper OWT3.

14. К Борен, Д. Хафмен, Поглощение и рассеяние света малыми частицами. (М.:Мир,1986).
    C. F. Bohren and D. R. Huffman, Absorption and Scattering of Light by Small Particles (Willey, New York, 1983).

15. М. Борн, Э Вольф. *Основы оптики*. (М.: Наука,1970).

M. Born and E. Wolf, Principles of Optics (Pergamon Press, Oxford, London, Edinburg, New York, Paris, Frankfurt, 1970).

16. B. Denker, B. Galagan, V. Osiko, I. Shulman, S. Sverchkov, E. Dianov, Applied Physics B *(DOI 10.1007/s00340-3406-2).*

17. Е. М. Дианов, С. В. Фирстов, В. Ф. Хопин, О. И. Медведков, А. Н. Гурьянов, И. А. Буфетов, *Квантовая электроника,* **39**, 299 (2009).

18. Л. И. Булатов, В. М. Машинский, В. В. Двойрин, Е. Ф. Кустов, Е. М. Дианов, Квантовая электроника, **40**, №2 153(2010);

19. Булатов Л И. *Абсорбционные и люминесцентные свойства висмутовых центров в алюмо- и фосфоросиликатных световодах. Автореферат дисс.кфм наук (Москва, МГУ, 2009)* ].

20. I. A.Bufetov, E. M. Dianov , Laser Phys. Lett. 1-18(2009)/ DOI 10.1002/lapl.200910025 **.**

21. Y. Fujimoto, M. Nakatsuka, Jpn. J. Appl. Phys. Part 1 **40**, L279 (2001).

22. Peng M.Y., Qiu J., Chen D., Meng X.,Yang I.,Jiang X., Zhu C., Opt. Lett. **29**,1998 (2004)

23. Peng M Y., Meng X., Qiu J., Zhao Q., Zhu C., Chem. Phys. Lett. **403**,410 (2005)

24. Peng M Y., Qiu J., Chen D., Meng X., Zhu C., Opt. Lett. **30**, 2433 (2005)

25. Ren J at all J., Mater. Res. **22**, 1574 (2007)

26. Meng X-W., Qiu J., Peng M Y., Chen D., Zhao Q., Zhu C *Opt. Express* **13**, 1628 (2005)

27. T. Suzuki T., Appl. Phys. Lett. **88**, 1912 (2006)

28. В. Е Оглуздин, XXIV Съезд по спектроскопии, Москва-Троицк, Россия, 28 февраля - 5 марта, 2010,тезисы докладов, том 2, стр.463.

Оглуздин Валерий Евгеньевич – Ogluzdin Valeriy E..

Кандидат физ.- мат. наук, старший научный сотрудник, ведущий инженер

Официальное название места работы.

Учреждение Российской академии наук Институт общей физики им. А. М. Прохорова РАН;   Prokhorovs general physics institute RAS;

119991 г. Москва, ул. Вавилова, д. 38.

Домашний адрес: 127422 г. Москва, ул. Астрадамская д. 4 кв. 28.

Телефон служебный: (499) 503 81 98.

Факс (499) 135 81 91

Электронная почта: ogluzdin@kapella.gpi.ru

Ogluzdin Valeriy E.